\documentclass{aa} 
\usepackage{natbib}
\usepackage{graphicx}
\usepackage{txfonts}
\usepackage{color}
\def\aap{A\&A}
\def\apjl{ApJ}

\def\apjs{ApJS}

\def\apjs{ApJ Supplements}

\def\apjl{ApJ Letters}

\def\aap{A\&A}

\begin{document}
   \title{The Cosmic Timeline Implied by the JWST Reionization Crisis}

   \author{F. Melia\thanks{John Woodruff Simpson Fellow}}

   \offprints{F. Melia}
\titlerunning{Cosmic Timeline in the Early Universe}
\authorrunning{Melia}

\institute{Department of Physics, The Applied Math Program, and Department of Astronomy, 
The University of Arizona, Tucson, Arizona 85721, USA; \email{fmelia@email.arizona.edu}}

   \date{Received May 31, 2024}

 
  \abstract
   {JWST's discovery of well-formed galaxies and supermassive black holes only a 
   few hundred Myr after the big bang, and the identification of polycyclic aromatic
   hydrocarbons (PAHs) at $z=6.71$, seriously challenge the timeline predicted by $\Lambda$CDM. 
   Moreover, the implied bright UV-irradiation of the early Universe suggests a reionization 
   history much too short to comply with the observed evolution of the hydrogen ionization 
   fraction, $x_{HII}(z)$.}
   {A recent analysis of reionization after JWST by \cite{Munoz:2024} has concluded
   that the $\Lambda$CDM timeline simply cannot accommodate the combined JWST-{\it Planck}
   observations even if exotic fixes are introduced to modify the standard reionization model.
   In this paper, we argue that this so-called `photon budget crisis' is more likely
   due to flaws in the cosmological model itself. We aim to re-analyze the data in the context
   of established astrophysics with $R_{\rm h}=ct$ as the background cosmology, an approach
   that has already been shown to mitigate the tension created by the too-early appearance 
   of galaxies, quasars and PAHs.}
   {We employ the standard reionization model using the JWST-measured UV luminosity
   function in the early Universe and the timeline and physical conditions in both 
   $\Lambda$CDM and $R_{\rm h}=ct$. The former has already been fully probed and discussed 
   by \cite{Munoz:2024}, and we here merely redo the calculation to ensure consistency with 
   their pioneering work. We then contrast the predicted reionization histories in these two 
   scenarios and compare them with the data.}
   {We confirm that the reionization history predicted by $\Lambda$CDM is in significant tension
   with the observations, and demonstrate that the latter are instead in excellent agreement with 
   the $R_{\rm h}=ct$ timeline.}
   {Together, the four anomalies uncovered by JWST, including the newly discovered reionization
   `crisis' discussed in this paper, provide strong evidence against the timeline 
   predicted by $\Lambda$CDM and in favor of the evolutionary history in $R_{\rm h}=ct$.}


   \maketitle
%

\section{Introduction}\label{intro}
Observational constraints indicate that the epoch of reionization
(EoR) probably began around redshift $z\sim 10-15$ and ended by
$z\sim 6$ \citep{McGreer:2015,Greig:2017,Sobacchi:2015,Mason:2019,Whitler:2020,Wang:2020,Nakane:2024}.
It is believed that Lyman continuum radiation from early galaxies
\citep{Robertson:2015} (with a possible contribution from emerging 
quasars; \citealt{Madau:2015}) reionized the expanding gas following 
the so-called dark ages that began soon after recombination at 
$t\sim 380\,000$ yr in the context of $\Lambda$CDM.

Ironically, the earliest attempts at understanding how reionization
came about struggled to reconcile an apparent paucity of ionizing
radiation at high redshifts with the timescale at which this
process took place, beginning with cosmic microwave background (CMB)
constraints from WMAP \citep{Komatsu:2011}, then improving with 
newer CMB data from {\it Planck} \citep{Planck:2016}, and even from
considerations \citep{MeliaFatuzzo:2016} of the alternative 
Friedmann-Lema\^itre-Robertson-Walker (FLRW) cosmology known as the 
$R_{\rm h}=ct$ universe \citep{Melia:2007,MeliaShevchuk:2012,Melia:2020}, 
the principal focus of this paper.

But JWST has completely reversed this picture, revealing a surprising
number of well-formed galaxies at $z$ approaching 14 and beyond, a mere
few hundred Myr after the big bang---much earlier than expected in the context
of $\Lambda$CDM. These first galaxies were prolific emitters of ionizing
photons which, as we shall confirm shortly, produced far too much UV irradiation
of the intergalactic medium (IGM) in the context of $\Lambda$CDM 
\citep{Munoz:2024} to allow reionization to extend over the observed 
redshift range $15\gtrsim z\gtrsim 6$.

The incorrect evolutionary history implied by this finding represents a fourth
strike delivered by JWST against the predictions of the standard 
model. Most glaringly, well-formed galaxies \citep{Melia:2023b,Lopez:2024}
and quasars \citep{Melia:2024b} have been discovered well before structure 
formation should have begun in the early Universe, followed by the
detection of polycyclic aromatic hydrocarbons (PAHs) at $z=6.71$
\citep{Witstok:2023,Melia:2024a}. These dust grains should have taken about
$1$ Gyr to form via the standard AGB channel, especially at low metallicities,
yet they appear a mere $\sim 500$ Myr after the first stars are supposed
to have formed in $\Lambda$CDM. 

In this paper, we shall argue that all four of these significant problems
consistently point to a single flaw in the standard model---its incorrect time 
versus redshift relation. We already know from earlier work \citep{Melia:2023b,Melia:2024a}
that the timeline predicted by the $R_{\rm h}=ct$ universe eliminates 
the first three of these anomalies. But the motivation for considering whether
this model might also mitigate the reionization crisis extends well beyond these
results. In recent years, we have carried out numerous comparative tests
between $R_{\rm h}=ct$ and $\Lambda$CDM, showing that the data tend to favor the
former over the latter with a likelihood $\sim 90\%$ versus $\sim 10\%$,
based on over 30 different kinds of observations, at both high and low
redshifts. A short summary of these publications may be found in 
Table~2 of \cite{Melia:2018e}, and a more complete description of the model's 
fundamental origin and scientific justification may be seen in \cite{Melia:2020}.

We shall begin by re-analyzing the standard reionization model in $\Lambda$CDM,
solely for the purpose of establishing consistency with the pioneering work
of \cite{Munoz:2024}. We assume flat $\Lambda$CDM with $h=0.7$, $\Omega_{\rm m}=
0.3$ and $\Omega_{\rm b}=0.0224/h^2$ to match previous fiducial models, and use 
AB magnitudes throughout \citep{Oke:1983}. As we shall see, our results fully 
confirm their conclusions, establishing a need to consider the alternative 
$R_{\rm h}=ct$ cosmology. We shall then repeat the analysis for the timeline 
and geometry in $R_{\rm h}=ct$, showing that both the reionization history and 
CMB optical depth in this model are well supported by the JWST and 
{\it Planck} data. We shall thereby conclude that the self-consistent 
resolution of all four anomalies uncovered by JWST via the replacement 
of $\Lambda$CDM with $R_{\rm h}=ct$ argues compellingly in favor of the 
latter as the correct cosmological model.

\section{Standard Reionization History}\label{expansion}
The three primary factors determining the conventional reionization history
include: (i) the average ionizing photon emissivity, (ii) the fraction 
$f_{\rm esc}$ of these photons escaping into the IGM, and (iii) the number
of recombinations per hydrogen atom. The first of these is given by the
abundance of galaxies at high $z$, expressed in terms of the UV luminosity function
(UVLF), $\Phi_{\rm UV}$, in units of the comoving number density of galaxies
per UV magnitude, $M_{\rm UV}$, times the ionizing efficiency $\xi_{\rm ion}$
of each galaxy. The new JWST observations show that early galaxies have 
higher ionizing efficiencies than previously assumed, with 
\begin{equation}
	\log_{10}(\xi_{\rm ion}/{\rm Hz\;erg}^{-1})\approx 25.8+0.11(M_{\rm UV}+17)
	+0.05(z-7)\,,\label{eq:xiion}
\end{equation}
versus $\sim 25.2$ \citep{Atek:2024,Simmonds:2024,Endsley:2023,Calabro:2024}.
Following \cite{Munoz:2024}, we shall conservatively cap $\xi_{\rm ion}$
at $z=9$ and $M_{\rm UV}=-16.5$ to avoid extrapolation.

The escape fraction in high-$z$ galaxies is not well motivated theoretically,
and difficult to measure while the IGM contains neutral hydrogen. Studies
of reionization-galaxy analogues at low $z$ have found a strong correlation
between $f_{\rm esc}$ and the UV slopes, $\beta_{\rm UV}$, however, which
provides a rough estimate \citep{Flury:2022,Chisholm:2022,Begley:2022}. 
Based on these studies, the slopes of JWST galaxies imply that 
$f_{\rm esc}\sim 5-15\%$ \citep{Mascia:2023,Lin:2024}.

We shall use the fit from \cite{Chisholm:2022},
\begin{equation}
	f_{\rm esc}=A_f\times 10^{\,b_f\beta_{\rm UV}}\;,\label{eq:fesc}
\end{equation}
where $A_f=1.3\times 10^{-4}$ and $b_f=-1.22$. This expression may be
combined with the $\beta_{\rm UV}-M_{\rm UV}$ measurements in
\cite{Zhao:2024}, incorporating data from both JWST and HST
\citep{Bouwens:2014,Topping:2022,Cullen:2023}, to estimate $f_{\rm esc}(M_{\rm UV})$.
The UV slopes are capped at $\beta_{\rm UV}=-2.7$ to avoid 
extrapolation, since these limits correspond to the bluest galaxies for
which Equation~(\ref{eq:fesc}) was calibrated.

The volume-averaged hydrogen ionized fraction, $x_{\rm HII}=1-x_{\rm HI}$,
where $x_{\rm HI}\equiv n_{\rm HI}/n_{\rm H}$ is the corresponding neutral
fraction, evolves according to the equation \citep{Madau:1999}
\begin{equation}
	\dot{x}_{\rm HII}={\dot{n}_{\rm ion}\over n_{\rm H}}-
	{x_{\rm HII}\over t_{\rm rec}}\;,\label{eq:Madau}
\end{equation}
where $\dot{n}_{\rm ion}$ is the rate of ionizing-photon production,
$n_{\rm H}=\rho_{\rm b}(1-Y_{\rm He})/m_{\rm H}c^2$, $Y_{\rm He}$ is the
helium mass fraction, $m_{\rm H}$ is the proton mass, and $\rho_{\rm b}$
is the baryon energy density.

The recombination timescale is given as \citep{Shull:2012}
\begin{equation}
	t_{\rm rec}=\left[C\alpha_{\rm B}(1+x_{\rm He})n_{\rm H}\right]^{-1}\;,\label{eq:trec}
\end{equation}
where $x_{\rm He}\equiv n_{\rm He}/n_{\rm H}\approx Y_{\rm He}/[4(1-Y_{\rm He})]$
is the helium fraction, $\alpha_{\rm B}$ is the case-B recombination coefficient,
and $C$ is the clumping factor. We shall follow previous work and set $C=3$ and
evaluate $\alpha_{\rm B}$ at $T=2\times 10^4$ K.

The ionizing-photon production rate may be expressed as
\begin{equation}
	\dot{n}_{\rm ion}=\int dM_{\rm UV}\,(1+z)^3\Phi_{\rm UV}\dot{N}_{\rm ion}f_{\rm esc}\;,
	\label{eq:dotnion}
\end{equation}
where the factor $(1+z)^3$ converts $\Phi_{\rm UV}$ into a density per
unit proper volume, and the integral extends down to a cutoff magnitude 
that is essentially a free parameter, though its impact on the results is 
small. The directly observed galaxies extend down to magnitude
$M_{\rm UV}\sim -15$. If one includes galaxies much fainter than this, the 
implied escape fraction $f_{\rm esc}$ would appear to be unrealistically small
\citep{Munoz:2024}. In addition, one expects a turnover due to feedback
\citep{Shapiro:2004}, which seems to be confirmed by HST observations,
with $M_{\rm UV}\sim 15$ \citep{Atek:2018}. In this paper, we shall 
therefore conservatively take the cutoff to be $M_{\rm UV}=-14.6$ 
in $R_{\rm h}=ct$, and assume the same value for $\Lambda$CDM as well. 
Our result for the standard model, summarized in Figure~\ref{fig1} below, 
is thus somewhat less extreme than that of \cite{Munoz:2024}, who instead
integrated Equation~(\ref{eq:dotnion}) down to $M_{\rm UV}=-13$. The UVLF 
is taken from the pre-JWST fit in \cite{Bouwens:2021} at $z\le 9$, and 
from \cite{Donnan:2024} at $z>9$. Lastly, $\dot{N}_{\rm ion}\equiv
L_{\rm UV}\,\xi_{\rm ion}$ is the ionizing-photon production rate per
galaxy, in terms of their UV luminosity $L_{\rm UV}$ and ionizing
efficiently $\xi_{\rm ion}$.

\begin{figure}
\centering
\includegraphics[width=3.5in]{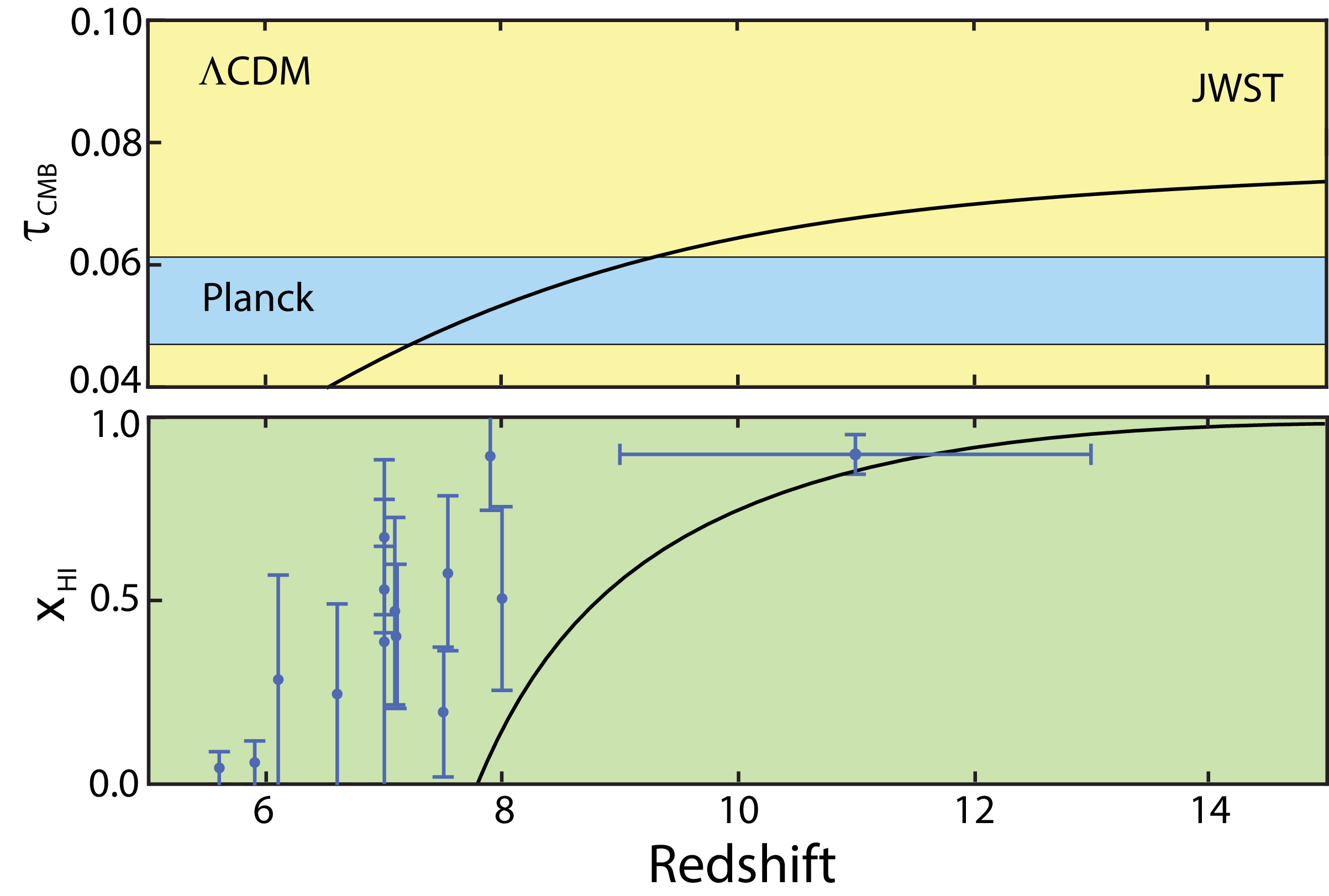}
\caption{Reionization history in $\Lambda$CDM based on new JWST
observations of structure formation in the early Universe. Lower panel:
the neutral fraction $x_{HI}$ as a function of redshift $z$, for a
cutoff $M_{UV}=-14.6$ and a JWST-calibrated ionizing efficiency,
$\xi_{\rm ion}$, and escape fraction ($f_{\rm esc}$) inferred from
low-$z$ analogues. The blue data points represent measured observational
constraints (see text for references). Upper panel: CMB optical depth,
$\tau_{\rm CMB}$, as a function of $z$. The yellow band 
represents the {\it Planck} measurement and its uncertainty
\citep{PlanckVI:2020}. The $\chi^2_{\rm dof}$ for
the $x_{\rm HI}$ fit is $4.36$, and $\tau_{\rm CMB}$ differs from
the {\it Planck} optimized value by $\sim 2\sigma$. The new JWST galaxy 
observations thus create significant tension for the standard model, both 
for the reionization history and the CMB optical depth.}
\label{fig1}
\end{figure}

Correspondingly, the reionization history is used to calculate the CMB optical
depth,
\begin{equation}
	\tau_{\rm CMB}=\int dR\, n_e\sigma_T\;,\label{eq:tauCMB}
\end{equation}
where $R$ is proper distance, $\sigma_T$ is the Thomson cross section, and
\begin{equation}
	n_e=\left(1+{\alpha Y_{\rm He}\over 4(1-Y_{\rm He})}\right)x_{\rm HII}\,n_{\rm H}
	\label{eq:ne}
\end{equation}
is the proper electron number density, with $\alpha=1$ for $z>4$ and
$\alpha=2$ for $z\le 4$. For simplicity, we assume that HeI reionization 
tracks HI, and that HeII reionization occurs at $z=4$.

The lower panel in Figure~\ref{fig1} shows the hydrogen neutral fraction as a
function of $z$ calculated from Equation~(\ref{eq:Madau}) with $x_{\rm HII}=0$ 
at $z=30$, assuming $\Lambda$CDM is the correct cosmology. This theoretically 
predicted curve is compared with the data (blue points), taken from 
\cite{McGreer:2015,Greig:2017,Sobacchi:2015,Mason:2019,Whitler:2020,Wang:2020,Nakane:2024}.
The upper panel shows the corresponding CMB optical depth, along with the 
{\it Planck} measurement (blue): $\tau_{\rm CMB}= 0.054\pm0.007$ \citep{PlanckVI:2020}.

We fully confirm the result, first presented by \cite{Munoz:2024}, that the new 
JWST observations clearly imply a reionization history in $\Lambda$CDM ending 
at $z\sim 9$. The disparity between the data and the predicted $x_{\rm HI}(z)$ 
and $\tau_{\rm CMB}(z)$ profiles creates significant tension for the 
standard model. In particular, note that the calculated CMB optical depth, 
$\tau_{\rm CMB}\sim 0.08$, misses the {\it Planck} measurement by almost $4\sigma$,
giving rise to the so-called `JWST reionization crisis.'

\cite{Munoz:2024} extended their analysis beyond this point, principally
to discuss all plausible physical mechanisms, theoretical and observational,
that might resolve this photon budget anomaly. The outcome of that work, however, 
is that all of the exotic fixes attempted thus far are in conflict with at least 
one observational constraint. There is yet no known solution to this emerging
problem with the standard model.

\section{Reionization History in $R_{\rm h}=ct$}\label{Rhct}
Let us now take the opposite approach, assume that the astrophysics of reionization 
is basically correct, and instead redo the calculation for $R_{\rm h}=ct$. There are 
several critical differences between the timeline and geometry of an $R_{\rm h}=ct$
universe and one driven by $\Lambda$CDM.

The first modification arises from the age-redshift relation. In $\Lambda$CDM,
we have
\begin{equation}
	t^{\Lambda{\rm CDM}}(z) = {1\over H_0}\int_z^\infty {du\over
	\sqrt{\Omega_{\rm m}(1+u)^3+\Omega_{\rm r}(1+u)^4+\Omega_\Lambda}}\label{eq:tLambda}
\end{equation}
where, in keeping with our assumption of a flat Universe and dark energy in the form
of a cosmological constant, we have simply $\Omega_\Lambda=1-\Omega_{\rm m}-\Omega_{\rm r}$.
The corresponding expression for $R_{\rm h}=ct$ is
\begin{equation}
	t^{R_{\rm h}}(z) = {1\over H_0(1+z)}\;.\label{eq:tRh}
\end{equation}
As demonstrated elsewhere, e.g., \cite{MeliaFatuzzo:2016}, these relations show that the
Universe is approximately the same age today in both models, but $t^{R_{\rm h}}$
is roughly twice $t^{\Lambda{\rm CDM}}$ at $z\gtrsim 6$. This is the reason, of course,
why the $R_{\rm h}=ct$ universe completely eliminates the `too early' galaxy and
supermassive black hole problems.

In addition, the two models differ significantly in terms of geometry. The 
differential comoving distance at $z$ in $\Lambda$CDM is given as
\begin{equation}
	dD_{\rm com}^{\Lambda{\rm CDM}}(z)={c\over H_0}{dz\over
        \sqrt{\Omega_{\rm m}(1+z)^3+\Omega_{\rm r}(1+z)^4+\Omega_\Lambda}}\,,
	\label{eq:dLambda}
\end{equation}
while in $R_{\rm h}=ct$ it is
\begin{equation}
	dD_{\rm com}^{R_{\rm h}}(z)={c\over H_0}{dz\over(1+z)}\;.\label{eq:dRh}
\end{equation}
Surprisingly, the two integrated comoving distances track each other rather closely 
up to $z\sim 8$, but then deviate considerably towards higher redshifts. For
example, the ratio of proper volumes at $z=15$ is $dV_{\rm com}^{\Lambda{\rm CDM}}/
dV_{\rm com}^{R_{\rm h}}\sim 0.1$.

\begin{figure}
\centering
\includegraphics[width=3.5in]{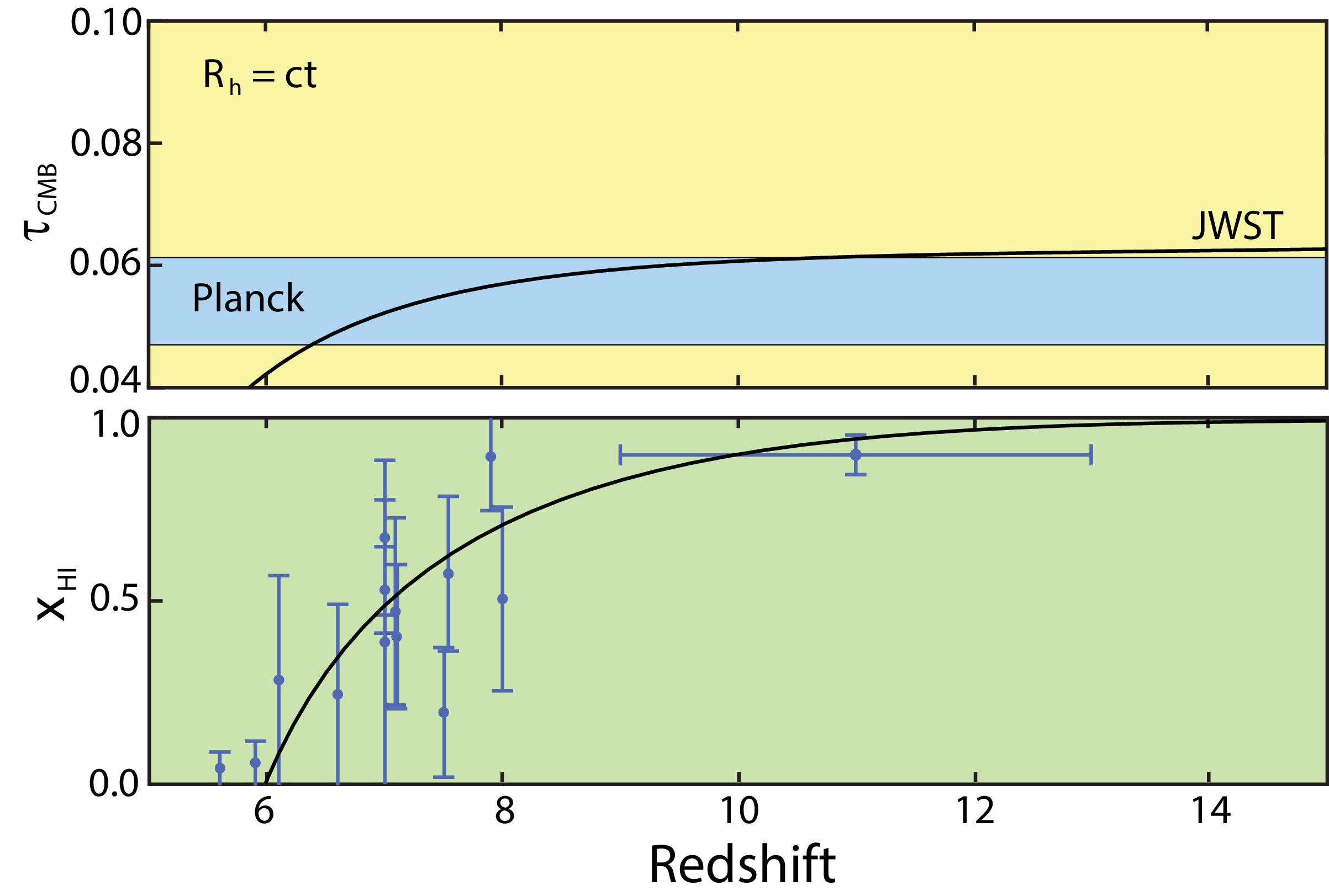}
\caption{Same as Fig.~\ref{fig1}, except now for the $R_{\rm h}=ct$
cosmology and a cutoff $M_{UV}=-14.6$. The $\chi^2_{\rm dof}$ for
the $x_{\rm HI}$ fit is $1.1$. The reionization history in this model is thus 
fully consistent with the UV irradiation of the early Universe by the newly 
discovered JWST galaxies, and the corresponding CMB optical depth falls within 
$\sim 1\sigma$ of its measured value even though, in principle, the optimized 
cosmological parameters in this model need not be identical to those inferred 
for $\Lambda$CDM.}
\label{fig2}
\end{figure}

Given these two principal differences between the two models, one should
also expect variations in the parameters, such as $H_0$ and $\Omega_{\rm b}$,
optimized to fit the full range of cosmological observations. But it turns
out these differences are usually rather small, mainly because it appears that
the formulation in $\Lambda$CDM, which was largely developed empirically,
produces an expansion history mimicking that of $R_{\rm h}=ct$. It is
beyond the scope of the present work to carry out such detailed comparisons,
which have actually already appeared in the literature, as one may see
from Table~2 of \cite{Melia:2018e}. For simplicity (and transparency) in 
this paper, we shall therefore adopt the $H_0$ and $\Omega_{\rm b}$
values optimized for $\Lambda$CDM.

Finally, to solve Equation~(\ref{eq:Madau}) starting with $x_{\rm HII}=0$
at $z=30$, we also need to recalibrate the UVLF $\Phi_{\rm UV}$ in 
Equation~(\ref{eq:dotnion}), assembled for $\Lambda$CDM, to make it 
compatible with the proper volume differences, $\propto (dD_{\rm com}^{\Lambda{\rm CDM}}
/dD_{\rm com}^{R_{\rm h}})^3$, between $\Lambda$CDM and $R_{\rm h}=ct$. 
The fits used for $\Phi_{\rm UV}(M_{\rm UV},z)$ are loosely 
based on the predictions of $\Lambda$CDM and $R_{\rm h}=ct$, but are 
largely empirical, as noted earlier, based on pre-JWST observations at 
$z\le 9$ \citep{Bouwens:2021} and JWST calibrations at $z>9$ \citep{Donnan:2024}. 
These expressions are inferred primarily from the observed number of 
galaxies as a function of $M_{\rm UV}$ and $z$, and the density is then 
determined using a conversion of measured redshift to model-dependent 
comoving volume. The data are model independent, as they depend only on $z$. 
Previous work had already confirmed that the halo mass function may be
formulated in a similar way in these two models \citep{Yennapureddy:2019},
the difference arising solely from a recalibration of the volumes. And
this is confirmed by a detailed comparison of the matter power spectra 
in $\Lambda$CDM and $R_{\rm h}=ct$, which fit the data equally well once 
the volume recalibration is taken into account \citep{Yennapureddy:2021}. 
Thus, the difference in $\Phi_{\rm UV}$ between the two models is due solely
to the difference in their predicted volumes, and for the analysis in this 
paper, it is therefore sufficient to merely recalibrate the UV luminosity
function used by \cite{Munoz:2024} via the change in comoving (or, equally, 
proper) volume from $\Lambda$CDM to $R_{\rm h}=ct$.

The calculated hydrogen neutral fraction in $R_{\rm h}=ct$ is shown as a 
function of $z$ in the lower panel of Figure~\ref{fig2}. The corresponding 
CMB optical depth in this model is shown in the upper panel. All the other 
features in this figure are identical to those in Figure~\ref{fig1}. 

Though we are not necessarily optimizing any model parameters
in this paper, it is nevertheless helpful to assess how well each of the
cosmologies fits the data in Figures~\ref{fig1} and \ref{fig2}. The 
$\chi^2_{\rm dof}$ of the fit based on the $\Lambda$CDM prediction 
(Fig.~\ref{fig1}) is $4.36$, while the calculated $\tau_{\rm CMB}$ is
$0.068$, which differs from the {\it Planck} value by about $2\sigma$. 
By comparison, $\chi^2_{\rm dof}$ for $R_{\rm h}=ct$ is 1.1 (Fig.~\ref{fig2}),
and $\tau_{\rm CMB}\approx 0.062$, which agrees with {\it Planck} to within
$\sim 1\sigma$. Were we to use these outcomes as a basis for 
head-to-head model selection using, e.g., the Aikake Information Criterion 
\citep{Akaike:1974,MeliaMaier:2013}, we would infer a relative
likelihood for $\Lambda$CDM of only $\sim 10^{-10}$, clearly implying
solid support for the $R_{\rm h}=ct$ cosmology by the 
reionization data.

\section{Discussion and conclusion}\label{discussion}
The mitigation of the JWST reionization crisis, from a comparison 
of Figures~\ref{fig1} and \ref{fig2}, is quite evident and robust. 
It is essential to understand that only the cutoff in 
$M_{\rm UV}$ for the integration in Equation~(\ref{eq:dotnion}) is 
somewhat undetermined, as described earlier. All one can say is that 
the cutoff must be $> -15$, and could have a value extending 
perhaps to $-12$ or $-13$, as assumed by \cite{Munoz:2024}. Of 
course, the larger this value is, the more serious the `crisis' 
becomes, because it enhances the UV irradiation of the IGM. We 
have here simply taken a conservative approach in using a cutoff 
as small as one could reasonably adopt based on the observations, 
in order to give $\Lambda$CDM the benefit of the doubt, so its
value was not fine-tuned, but merely assumed to be somewhat below 
the observational constraints, i.e., $-14.6$. All the other factors 
used in this analysis, including $f_{\rm esc}$, $\xi_{\rm ion}$, 
$\Phi_{\rm UV}$ and so forth, were strictly fixed by the previous
analyses.

It is thus rather easy to understand what `solves' the problem. By far, the
most significant difference between the calculation of $x_{\rm HII}(z)$ 
in $\Lambda$CDM and $R_{\rm h}=ct$ is the impact on the UVLF $\Phi_{\rm UV}$ 
of the proper volume. The Universe was larger at $z\gtrsim 10$ in $R_{\rm h}=ct$
than it would have been in $\Lambda$CDM. As a result, $\Phi_{\rm UV}(z)$ is smaller
in the former than in the latter for the same number of galaxies observed at $z$.
The UV-irradiation of the IGM was therefore less severe in $R_{\rm h}=ct$,
which lowered the reionization rate and thereby delayed the end of the
reionization epoch (to $z\sim 6$).

We have now witnessed at least four major discoveries by JWST in the early Universe 
that have created significant tension for $\Lambda$CDM. And
while exotic fixes may be introduced to fix each of them individually, e.g., by
proposing that supermassive black holes were created as $\sim 10^5\;M_\odot$
seeds \citep{Yoo:2004,Latif:2013,Alexander:2014}---a process that has never 
been seen anywhere in the cosmos---eventually one must acknowledge the fact that
a scenario in which a single modification simultaneously eliminates all of the
anomalies ought to be given serious consideration as being the correct approach.
Replacing $\Lambda$CDM with $R_{\rm h}=ct$ as the background cosmology
certainly does this---and rather impressively---as we have seen in this paper.

Looking forward, the prospect of $R_{\rm h}=ct$ being the correct cosmology
is quite alluring because it solves many, if not all, of the paradoxes and
conflicts accumulated by the standard model. For example, it eliminates all
horizon problems, including those associated with the CMB temperature and
the Electroweak phase transition \citep{Melia:2013c,Melia:2018e}. And it 
eliminates the cosmic entropy anomaly \citep{Melia:2021d}, the monopole 
problem \citep{Melia:2023f}, and inflation's violation of the strong energy
condition in general relativity \citep{Melia:2023e}, among many others.

Finally, recent fundamental work with the FLRW metric itself has demonstrated
that the overly simplified FLRW ansatz, with a lapse function $g_{00}=1$, may 
in fact be valid only for the zero active mass equation of state, $\rho+3p=0$,
in terms of the total energy density $\rho$ and pressure $p$, which produces
the constant expansion rate in $R_{\rm h}=ct$ \citep{Melia:2022b}. In other 
words, $R_{\rm h}=ct$ may be the only valid FLRW solution consistent with the 
foundational symmetries in the Cosmological principle.


\bibliographystyle{aa}
\bibliography{ms}

\end{document}